\documentstyle[12pt,aasms4]{article}

\begin{document}

\title{The highly polarized  open cluster Trumpler 27\altaffilmark{1}}

\author{Carlos Feinstein\altaffilmark{2}, Gustavo Baume\altaffilmark{3},   Ruben
Vazquez\altaffilmark{2}, Virpi Niemela\altaffilmark{4}}

\altaffiltext{1}{Based on observations obtanined at  Complejo Astron\'omico El
Leoncito (CASLEO), operated under agreement between CONICET and the National Universities 
of La Plata, C\'ordoba, and San Juan, Argentina.}

\altaffiltext{3}{On a fellowship from CONICET, Argentina.}
\altaffiltext{2}{Member of Carrera del Investigador Cient\'{\i}fico, CONICET, 
Argentina.}
\altaffiltext{4}{Member of Carrera del Investigador Cient\'{\i}fico, CIC, Prov. 
Buenos Aires, Argentina.}

\affil{Observatorio Astron\'omico, Paseo del Bosque, B1900FWA La Plata, Argentina } 

\and

\author{Miguel Angel Cerruti\altaffilmark{2}}

\affil{Instituto de Astronom\'{\i}a y F\'{\i}sica del Espacio, CC 67, Sucursal 28, 1428 Buenos Aires, Argentina }


\begin{abstract} 

We have  carried out multicolor linear polarimetry (UBVRI) of the brightest stars in the area of the 
open cluster Trumpler 27. Our data  show a high level of polarization in the stellar 
light  with a considerable dispersion, from $P = 4\%$ to $P = 9.5\%$. 
The polarization vectors of the cluster members appear to be aligned.
Foreground polarization was estimated from the data of some non-member objects, for which 
two different components were resolved: the first one associated with a dust cloud 
close to the Sun producing   $P_{\lambda max}=1.3\%$ and $\theta=146$ degrees, and a 
second component, the main source of polarization for the cluster members, originated in 
another dust cloud, 
which polarizes the light in the direction of $\theta= 29.5$ degrees. 
From a detailed analysis, we found that the 
two components have associated values $E_{B-V} < 0.45 $ for the first one,
and  $E_{B-V} > 0.75$ for the other.
Due the  difference in the 
orientation of both polarization vectors, almost 90 degrees (180 degrees at the Stokes representation), 
the first cloud ($\theta \sim 146$ degrees)  depolarize  the light strongly
polarized by the second one ($\theta \sim 29.5$ degrees).

\end{abstract}


\section{Introduction}

Trumpler 27  ({\it l} =355, {\it b} = -0.7), also known as C1732-334, is a heavily reddened open cluster 
located  approximately in the direction to the galactic center. 
This cluster deserved a major attention in the 
past because some of its brightest members are stars that play a dominant role when 
trying to interpret the different evolutionary phases of the brightest massive stars. 
Tr 27-102 is a long period Cepheid (Van Genderen et al. 1978, Bakker et al. 1981). 
Tr 27-28 and Tr 27-105 are two known Wolf-Rayet stars (WR 95 and WR 98 respectively) 
in the catalogue of galactic WR stars (van der Hucht et al. 1981). 
The first of them has a very large infrared excess due to thermal re-emission 
of the stellar UV radiation by dust grains (Th\'e et al. 1980).

The literature reports three previous studies of the stellar population in Tr 27 with the main 
interest to determine the distance to the cluster, namely  Th\'e and Stokes (1970); Moffat, FitzGerald \& Jackson (1977,
hereafter MFJ77); Bakker \& Th\'e, (1983, hereafter BT83). This last study determined a distance of
$1.65 \pm 0.25\ kpc$ using five color (WULBV) Walraven photometry, locating the Tr 27 cluster in the 
Sagittarius arm. MFJ77 found Tr 27  to be $6\ 10^{6}$  year old cluster containing 
eight supergiants stars,  six are blue (\#2,8,23,43,46,46a), one is yellow (\#102) and one is red (\#1).

MFJ77  also argued due to the lack of detected H$\alpha$ emission or a reflection nebula that 
the  layer of dust causing the strong reddening of  the cluster ($E_{(B-V)} \sim 1.25$) is in front, 
but not associated with  Tr 27. 
BT83 also found a normal extinction law  analyzing 3 stars observed in several wavelengths
and suggested that the radiation of these stars is already diluted when it 
reaches the dark cloud.



In order to understand the physical properties of the interstellar medium (ISM) towards 
Trumpler 27, we have carried out linear polarization observations of its brightest 
stars. In the next sections we will discuss the observational procedures, the data 
calibration and the results in terms of both individual stars and the whole cluster.

\section{Observations and data reduction}

Data of linear polarimetry were obtained during three observing runs at the Complejo Astron\'omico 
El Leoncito (CASLEO) San Juan, Argentina, using two different photopolarimeters 
attached to the 2.15 m telescope. 
The first observations were performed using the Vatican Polarimeter (VATPOL) 
during June 11-15, 1991, while the 
rest were carried out with the Torino Five Channel Photopolarimeter in May 30-June 3, 1995 
and June 30, 1997. Most of the stars were observed through the Johnson broad band UBVRI filters ($\lambda_{Ueff}=0.360 \ \mu m$,
$\lambda_{Beff}=0.440\ \mu m$, $\lambda_{Veff}=0.530\ \mu m$, $\lambda_{Reff}=0.690\ \mu m$, $\lambda_{Ieff}=0.830\ \mu m$), but a few fainter ones were observed in white light, i.e. without any filter (VATPOL run only). Several polarization standard stars (for angle and zero point) were also measured for calibration purposes. 
In order to verify the lack of systematic differences between the observations performed with  
both instruments, we show in Fig 1a,b the results obtained for a set of stars observed 
with both polarimeters. Notice that the straight lines in these figures are not least 
squares fits 
but the 45 degrees slopes drawn as a reference. No systematic 
difference was detected between the VATPOL and the Torino Five Channel Polarimeter data, as can
be seen from the excellent agreement, as well for the values of the polarization vectors, 
as for their respective angles.

Our results are listed in table 1 which shows, in self explanatory format, the stellar 
identification as given by 
MJF77, the average of the percentage of polarization (P), and the position 
angle of the electric vector ($\theta$) observed through  each filter, with their respective mean 
errors.

Several stars in the area have been observed by the Hipparcos astrometric satellite and are included in the 
Hipparcos/Tycho Catalogue Data (\#1,2,16,23,24,43,102,103 and 104). However, all these stars 
are so far away from the Sun that no useful parallax measures could be obtained.  Although \#24 is 
 a non-member star in front of the cluster, it too has a meaningless parallax in
Hipparcos data.

\section{Results}

We want to emphasize, as seen from data in Table 1, the presence of cluster stars having 
linear polarization values reaching 9\%. As far as we know, such  values are not very 
often measured in an open cluster. The only case reported in the literature having measurements
of polarization greater than Tr 27, is M17 where Schultz et al. (1981) found values in excess of 20\% for
some stars.

In figure 2 we show  the sky projection of the V band polarization for the observed stars in Tr 27. 
As a reference the dashed line is the galactic parallel $b=-0^{o}.8$ . Note the alignment of the
polarization vectors with the galactic plane.
An evident feature in this figure is also that some stars with low polarization (near $\sim$ 1\%) 
do not follow the general trend shown by most of the stars in Tr 27.
These stars (\#4,6,22,24 and 26) were  considered non-members in earlier investigations and they are very 
probably located in front of the cluster and the dark cloud  near the cluster 
(MFJ77, BT83).

Figure 3a presents the histogram of the angle distribution  of the polarization vectors for the 
observed stars. The non-member stars are easily detected due to their different angle, 
appearing as an isolated group at  $\sim$ 146 degrees. These objects can be used  to estimate
the interstellar polarization component (IP) in front of  Tr 27 and to subtract 
its contribution from the 
measures of the cluster stars. Averaging the value for non-members stars, we find that the IP component is 
 $P_{V}=1.32 \pm 0.02 \% $ and $\theta_{V}=146.6 \pm 5$ degrees. 
This angle is not  aligned with the galactic plane, but appears almost perpendicular, 
showing a perturbation in the local magnetic field of the Galaxy 
on the line of sight to the Tr 27 cluster.
As we do not know the distances to these foreground stars we can not estimate the extent over 
which the dust is aligned in this angle. 
From the data in the Catalogue of linear polarization of Axon and Ellis (1976), the angle of the polarization vector in this direction seems to show  a complex pattern not associated with the direction of the galactic plane in a large range of distances. In the photometry of MFJ77, most of these non-member objects have colors compatible with  low mass main sequence stars.

Figure 3b shows the plot of the observed linear polarization vector $P_{V}$ vs the polarization angle. In this figure the segregation between the members of Tr 27 and non-members becomes more obvious, as the cluster stars appear much more polarized.  
It is interesting to note that the non-member stars 4 and 6 ($\sim$ 2.5' to the W from stars \#22,24,26) are $\sim$ 0.2\% more polarized than the other non-members (also the angle is 10 degrees lower), which means that probably the IP component is not fixed at the field and probably has a gradient (Fig 3b) like the one found in the Carina Nebula  by Marraco et al. (1993). However, with only a  few objects
to determine the IP, a selection in distance can not be ruled out.

In Fig. 3a,  the stars observed in Tr 27 cluster  appear very close to $\sim$ 30 degrees in 
polarization angle, 
and the data can be  easily fitted with a Gaussian distribution.
The dispersion of this fit  is $\sim$ 10 degrees, which is in the range of the values found
by Waldhausen et al. (1999) for the open clusters NGC 6167, NGC 6193 and NGC 6204 in Ara OB1.   
 One star  in Tr 27 (\#103)  seems to be  less 
polarized than the others. Star \#103 is an isolated object 
located  1\'  to the West where the ISM has different properties. This star 
was not considered for the fitting of the Gaussian distribution.

We note that the direction of pointing of the  foreground component (IP) is  146.6 degrees, which is at near 90 degrees to the 
cluster component ($\sim$ 29.5 degrees), so that the total effect is to depolarize the light crossing through the ISM that produces the IP. Thus, the polarization percentages corrected for 
IP would be even  larger than the observed ones.

BT83 suggested that the star \#105 (WR 98) is a background object. However,
our data indicate that the light of this star has polarimetric properties ($P_{V}=5.2 \pm 0.04\%, \ \theta= 49.1 \pm 0.2$ degress) similar to the cluster members, because of the high polarization and the polarization angle average ($\theta \sim 30 \pm 10$ degress).
 As this star is a Wolf-Rayet type binary system (Niemela 1999) an intrinsec polarization componentmay be  expected, 
but it would be hard to distinguish from the high value of the polarization of the ISM.  
We consider WR98 to be a prabable member of the Tr 27 cluster.

\section{Analysis and Discussion}

To analyze the data, the polarimetric observations were fitted using  the Serkowski law of interstellar 
polarization  (Serkowski 1973). This is:

$P_{\lambda}/P_{\lambda max}=e^{-Kln^2(\lambda_{max}/\lambda)} \  \  \ \ [1]$

If the polarization is produced by aligned interstellar dust particles, the
observed data (in terms of wavelength, UBVRI) will follow [1], where each star is characterized by a $P_{\lambda max}$ and a $\lambda_{max}$.

Adopting $K = 1.66  \lambda_{max}+ 0.01$  (Whittet et al., 1992) , we fitted our observations and computed the $\sigma_{1}$ parameter
(the unit weight error of the fit) in order to quantify the departure of our
data from the ``theoretical curve'' of  Serkowski's law. A $\sigma_{1}$   larger than 1.5   is considered as an indication of the presence of 
a component of intrinsic stellar polarization. Another criterion of intrinsic stellar polarization is to compute the dispersion of position angle 
for each star normalized by the average of the position angle errors ($\bar{\epsilon}$).

The  $\lambda_{max}$ values can also be used to test the origin of the polarization. In fact, since the average value  of $\lambda_{max}$ 
for the interstellar medium is 0.545 $\mu m$ (Serkowski et al. 1975), objects showing  $\lambda_{max}$
rather lower than this value are also candidates for having an intrinsic component of polarization (e.g. Orsatti et al. 1998).
The values that we have obtained for $P$, the $\sigma_{1}$ parameter, $\lambda_{max}$, and $\bar{\epsilon}$ together with the  identification of stars  
are listed in Table 2.

Figure 4 shows the observed $P$ and $\theta$ of those stars which are the candidates for having an 
intrinsic 
component of polarization. For purposes of comparison, the best fit to a Serkowski's law to these observations is also plot as a continuous line. Stars \#1 and \#2 (spectral types M0Ia and O9Ia, respectively, MFJ77) have a large departure of the Serkowski's relation for the data at the wavelengths of  the I filter. Star \#1 also shows a significant  rotation in the position angle of the polarization vector. Both cases are noticeable despite the high polarization vector component added by the the dust, meaning a considerable intrinsic component of polarization. Stars \#25 and \#106 do not fit the Serkowski curve at U and I filters wavelengths.

The histogram of all the observed  $\lambda_{max}$ shown in Fig. 5 confirms that the most probable value of 
$\lambda_{max}$ for stars in Tr27 is around 0.55 m$\mu$, which is 
the same value found by Serkowski et al (1975) as the average value for the ISM.

The observed polarization $P(\%)$ and $\theta$ of the non-member stars show that there exist dust particles in a different  alignment than
those  observed in the  highly polarized  light of the Tr 27 cluster stars. 
We ill try to estimate  the effect of this layer of dust particles, characterizing its properties
(P(\%),$\theta$, $E_{B-V}$, etc).
The polarization vector and the angle of orientation can be obtained
from the non-members stars \#4,6,20,22 and 24, which give by fitting a Serkowski law to each one of these stars
 an average  of $P_{\lambda max}=1.35$ and $\theta=146.6$. 

Figure 6 is the plot of the $P_{\lambda max}$ (corrected by the IP vector) vs $E_{B-V}$ (data from MFJ77), but not 
corrected for $E_{B-V}$ of the IP, because the  value  is unknown. 
The dashed  line is the empirical upper limit relation for the interstellar polarization, 
$P_{\lambda max} = R\ A_{v} \sim 9 E_{B-V}$  (for normal dust, $R=3.2$), Serkowski et al. (1975).
If we consider that the  star \#23 is nearly the case of the maximum observed efficiency for the dust to polarize light, we can shift the dashed line over the $E_{B-V}$ axis  up to the location of star \#23 (the solid line in Fig. 6). This displacement ($\Delta E_{B-V} = 0.45 $) then represents the maximum $E_{B-V}$ allowed for the IP component ($E_{B-V \ IP}$). Note, that if the slope of the relation $P_{\lambda max}$ vs $E_{B-V}$ is lower than 9, as it probably  is, the maximum value of $E_{B-V}$ for the IP component will decrease. Therefore, we can conclude that the IP component at least must have an $E_{B-V \ IP} \leq 0.45 $

MFJ77 and BT83 have confirmed  that the total mean color excess towards Trumpler 27 is $E_{B-V}=1.2$. As we found that  $E_{B-V \ IP} \leq 0.45 $, the cloud that causes the high polarization must be associated with $E_{B-V} > 0.75$ (second component).  
Also both above mentioned papers  argue  that this last layer of dust (the second component) must be in front but not associated with the Tr 27 cluster due to the lack of H$\alpha$ emission or a reflection nebula (MFJ77); and since the extinction law appears normal for 3 luminous stars the radiation of these stars is already diluted when it reaches the dark cloud (BT83). 
On the other hand, in both papers their two-color plots show  a wide dispersion and not a sharp sequence, implying that  some of this dust may be intracluster or just in front of the cluster, but with a  density distribution that appears  very non-homogeneous.   

Using the  MFJ77 data for the non-member stars, and considering the value  $E_{B-V \ IP} < 0.45 $, we find that some of these objects are consistent with solar type main sequence stars. To achieve  the average polarization  $P_{\lambda max}=1.35$ for normal efficiency of the polarizing properties of the dust ($P_{\lambda max}/ E_{B-V} \sim 5 $, Serkowski et al., 1975) the $E_{B-V \ IP}$ can not be larger than $E_{B-V \ IP} \sim 0.3 $,  which is still in the range of $E_{B-V \ IP} \leq 0.45 $.


Excluding the cluster stars with the highest 
polarization percentage, namely \#23,25 and 27 and also the star \#103 with the lower polarization, the rest of the Tr 27 stars seems to be aligned in Fig. 6. The slope of this alignment is the polarization efficiency of the ISM. The fit for these stars gives $P_{\lambda max}/E_{B-V}= 4.6 \pm 0.12$, which is similar to the value of  $P_{\lambda max}/E_{B-V}=5$ mentioned above as the canonical value for ISM. 

This picture of two dust components agrees with the behavior of the interstellar absorption in this direction of the galaxy according the study of the spatial distribution of the  interstellar dust by Neckel \& Klare (1980). Their work shows a strong jump in absorption from $\ 1\ mag$ to $4\ mag $ at about 1 Kpc.

\section{Conclusions}

We have observed linear polarization ($P$ and $\theta$) for a sample of stars in the open cluster Trumpler 27, and also  a few
non-member objects. Our observations indicate that the cluster members show a percentage of polarization up to 9\%, an unusually high value for an open clusters. The dispersion of polarization values goes from 4 to 9\%, while the average orientation of polarization is 29.5 degrees, with a dispersion of 10 degrees approximately.

Using the non-member stars, located in front of the cluster, we could identify a first component of interstellar polarization (IP),
which is characterize by  $P_{V}=1.32 \pm 0.02 $ and a mean polarization angle of $\theta_{V}=146.6 \pm 5$ degrees. The second component that accounts for the bulk of the high polarization properties in Trumpler 27, has a mean polarization angle of $\theta_{V}=29.5$ degrees, oriented along the galactic plane.

From the analysis of the relation between extinction and polarization it is possible to estimate the extinction related with each dust cloud component. For the group of non-member stars, we found  $E_{B-V}<0.45$, a value which is in good
agreement with previous photometric values. The second component must have an $E_{B-V}>0.8$ to account for the previous 
photometry (MFJ77 , BT83).

\acknowledgments
We wish to acknowledge the technical support of CASLEO during our observing runs. We thank the Vatican Observatory for the loan of their polarimeter. Also we want to acknowledge useful discussions with Ana M. Orsatti, Nidia Morrell and Hugo Marraco, which are greatly appreciated.

\newpage


\begin{figure}\caption{Comparison of the data for the stars observed with both polarimeters  }\vspace*{ 0.25 truein}\hspace*{ 0.25 truein} \epsscale{0.7} \plotone{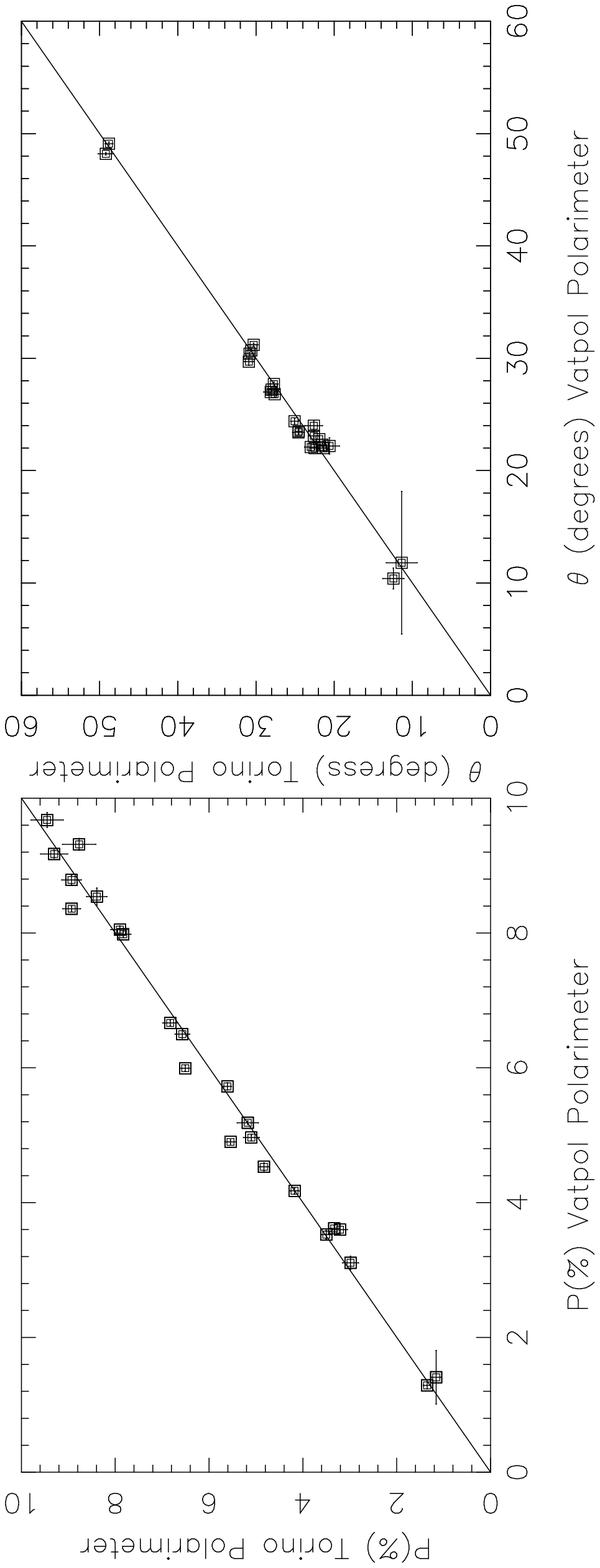}\end{figure}                                                 

\begin{figure}\caption{       Projection of the polarization vectors (Johnson V filter) over the sky. The dotted line is the galactic parallel $b=-0^{o}.8$ } \vspace*{ 0.25 truein}\hspace*{ 0.25 truein}\epsscale{0.7} \plotone{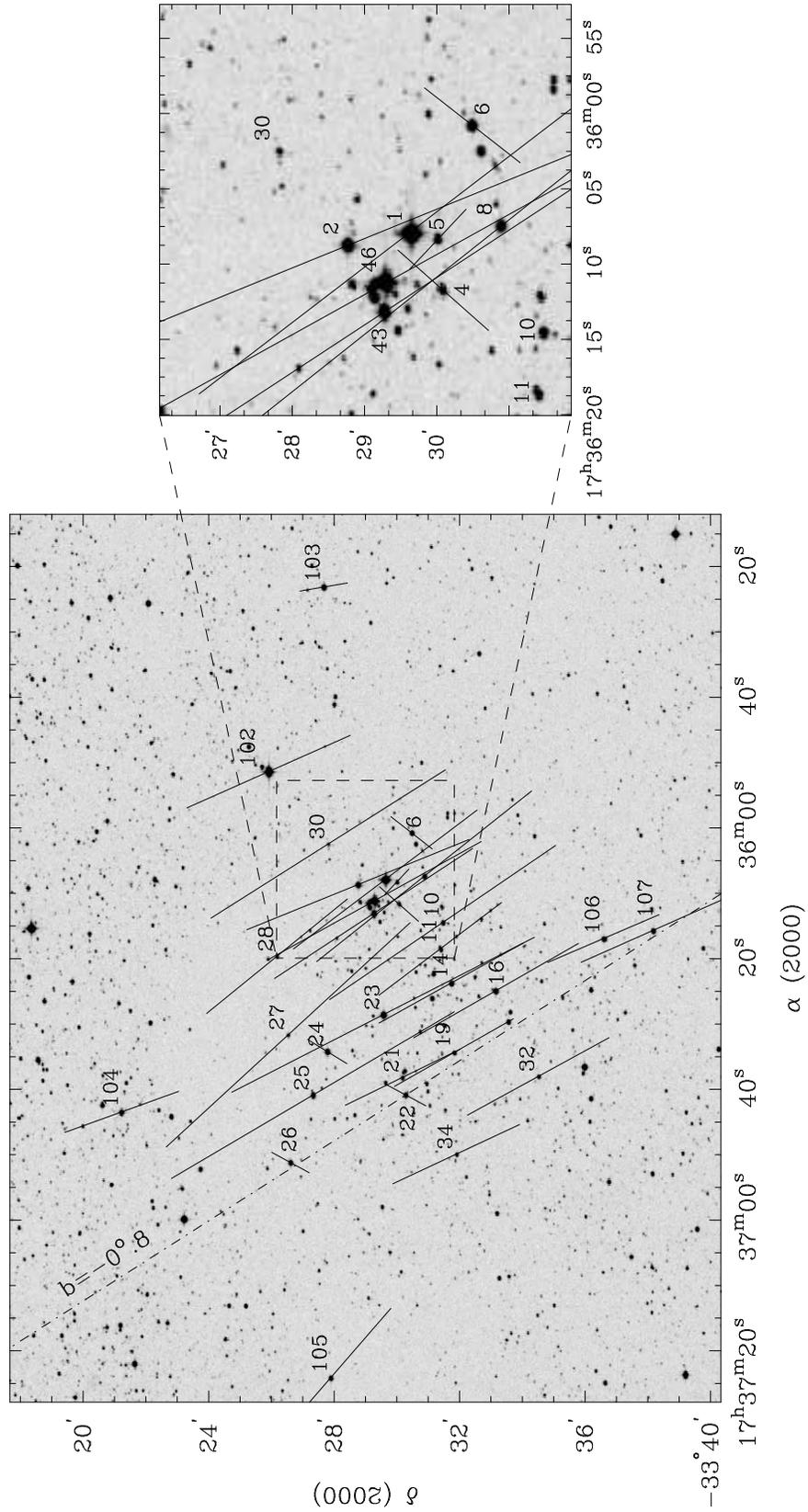}\end{figure}

\begin{figure}\caption{Top: Histogram of the polarization angle ($\theta$) for the observed stars. Shadowed bars are for the Tr 27 cluster stars and white bars are for the foreground stars.  The continuous line is the  Gaussian fit to the data for members of Tr 27. Bottom: Polarization percentage of the stellar flux versus the polarization angle for each star. Note how the cluster stars and the non-member objects are segregated.}\vspace*{ 0.25 truein}\hspace*{ 0.25 truein}\epsscale{0.7} \plotone{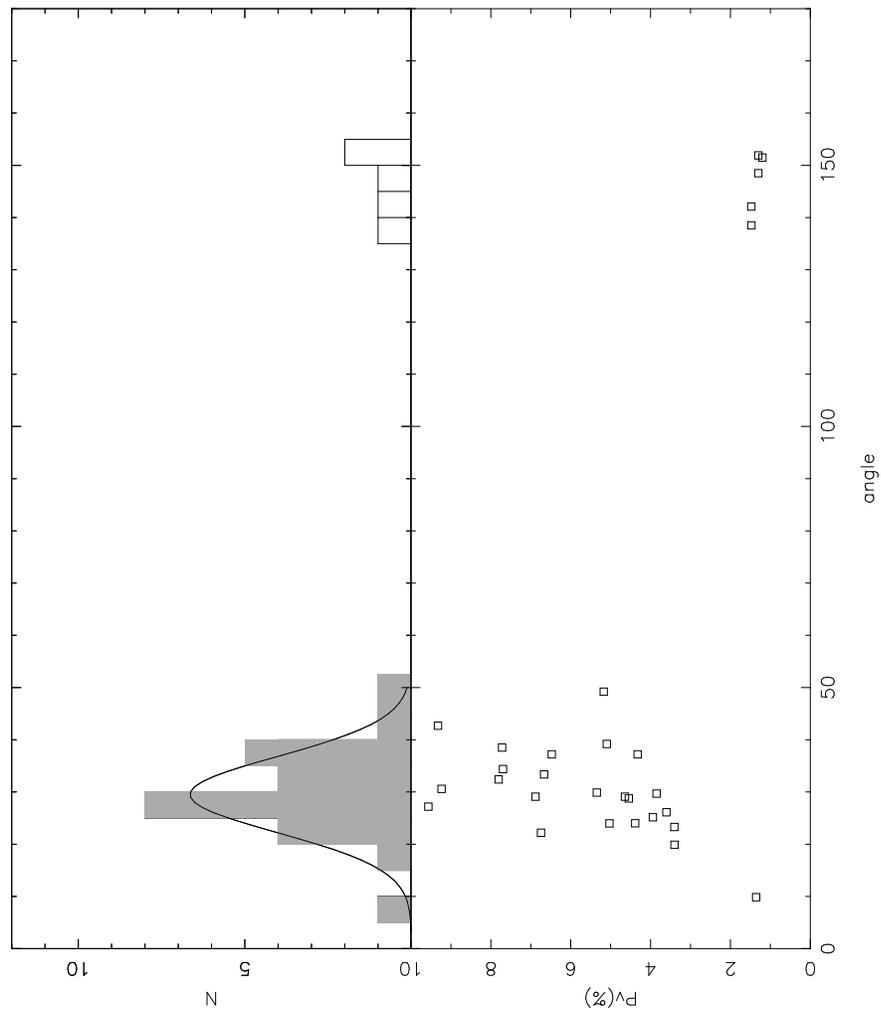}\end{figure}

\begin{figure}\caption{Plot of the observed data for objects showing large departures from the Serkowski law, the solid line is the best fit.}\vspace*{ 0.25 truein}\hspace*{ 0.25 truein}\epsscale{0.7} \plotone{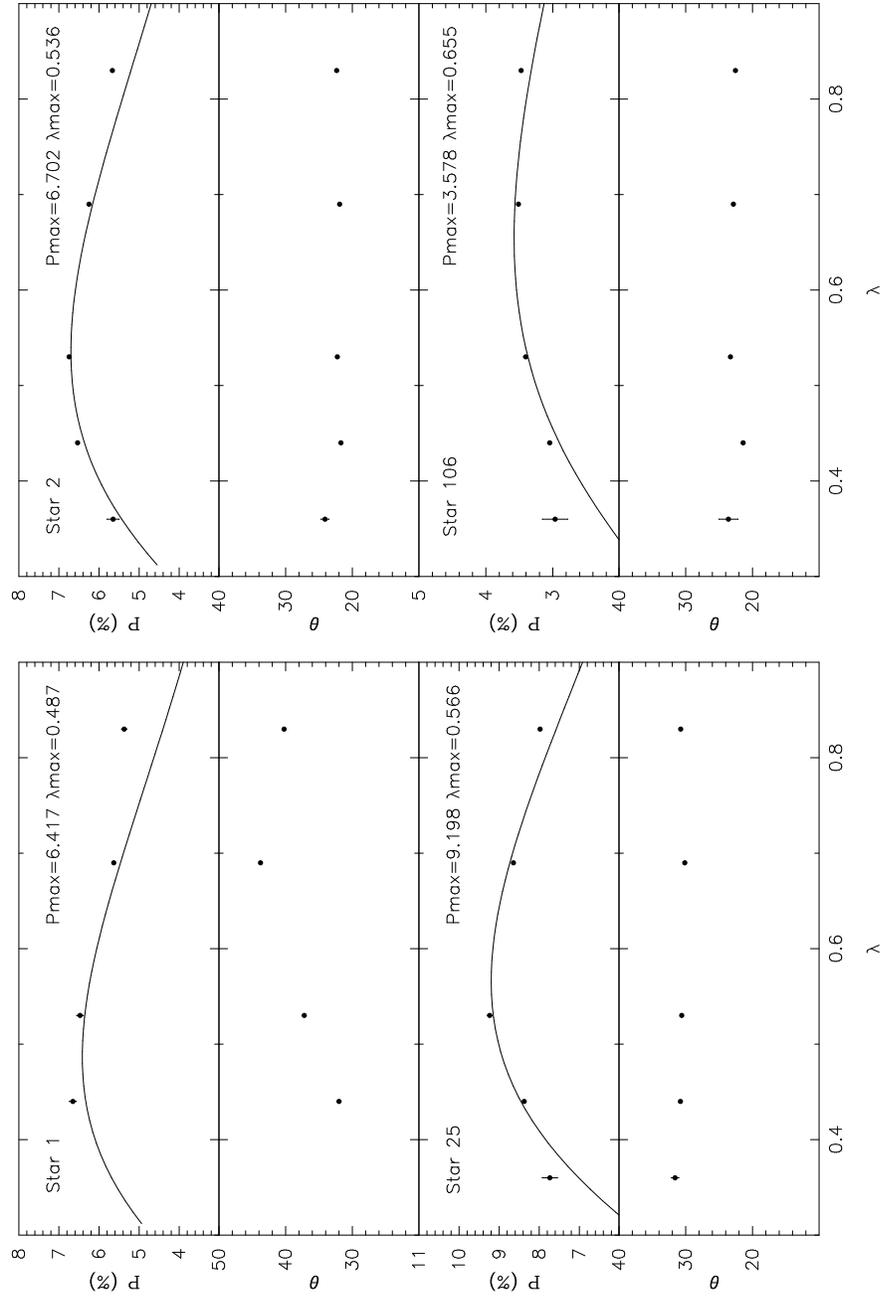}\end{figure}

\begin{figure}\caption{Histogram of the parameter $\lambda_{max}$ obtained fitting  the Serkowski law to the observations. }\vspace*{ 0.25 truein}\hspace*{ 0.25 truein}\epsscale{0.7} \plotone{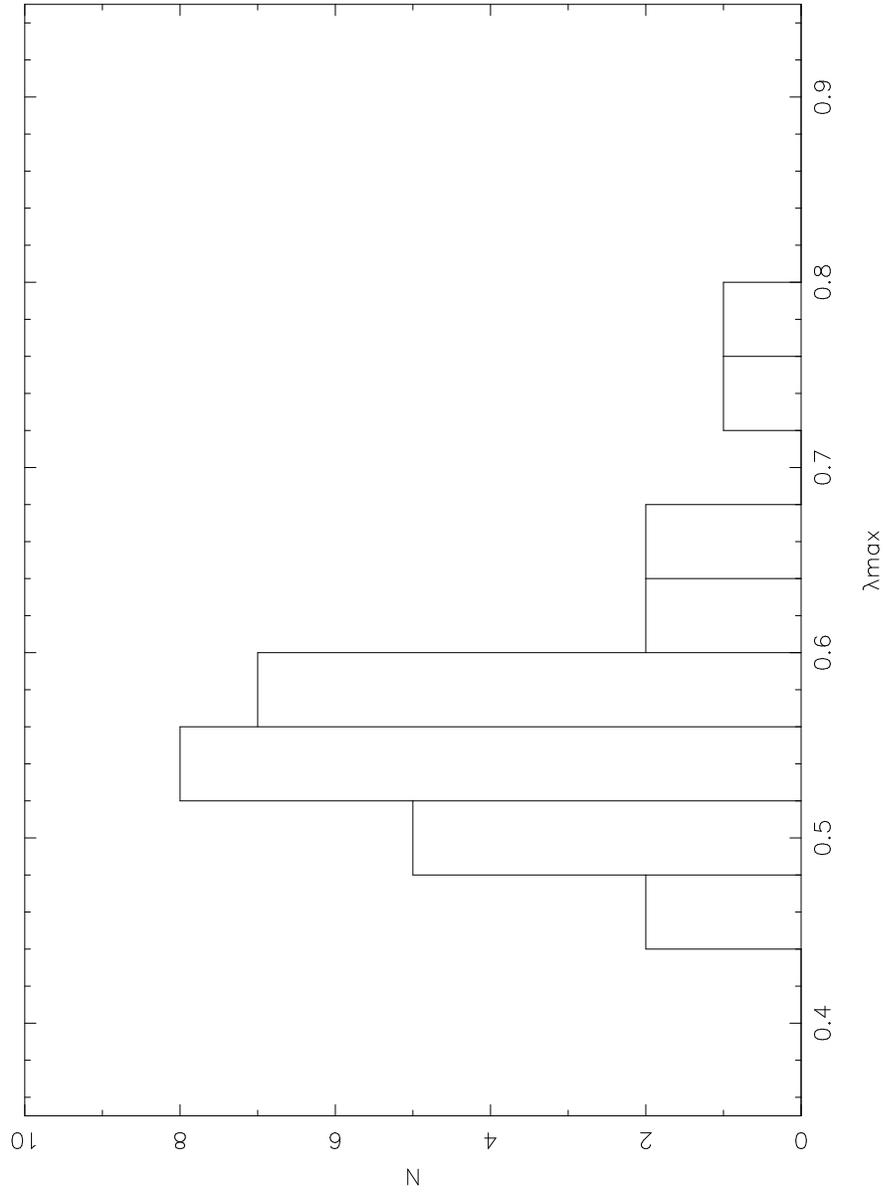}\end{figure}

\begin{figure}\caption{$P_{\lambda max}$ vs $E_{B-V}$ . The dashed line is  $P_{\lambda max} = \ 9 E_{B-V}$, and the solid line is a parallel for an IP component of $E_{B-V}=0.45$}\vspace*{ 0.25 truein}\hspace*{ 0.25 truein}\epsscale{0.7} \plotone{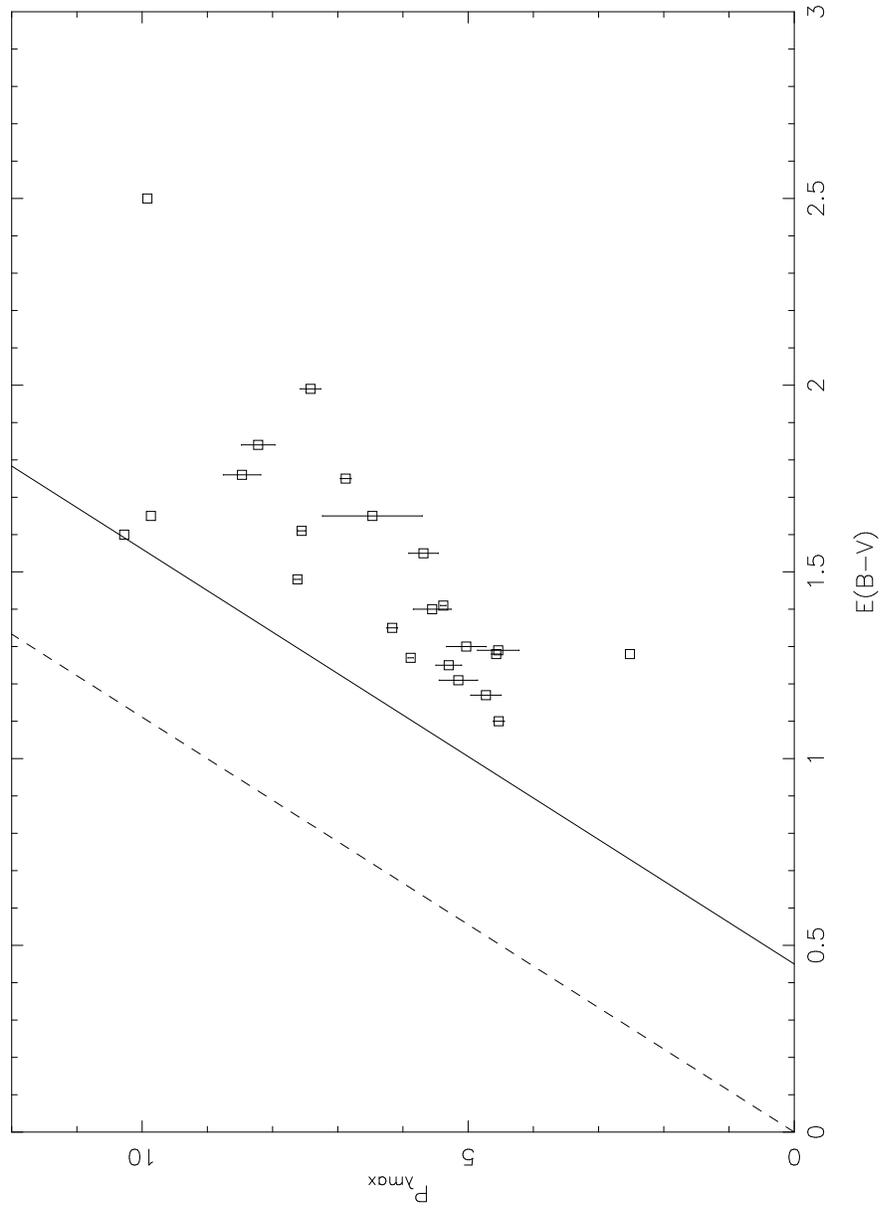}\end{figure}

\input{Feinstein.tab1.tex}
\begin{planotable}{ccccc}

\addtocounter{table}{+1} 
\tablecaption {Parameters of the Serkowski fit to the linear polarization data for stars in Tr 27}
\tablehead{
  \colhead{Stellar} &
  \colhead{ $P_{max} \pm \epsilon_{P}$} &
  \colhead{ $\sigma_{1}$ } &
  \colhead{$\lambda_{max} \pm \epsilon_{\lambda_{max}}$ }&
 \colhead{ $\bar{\epsilon}$ } \\
     \colhead {Identification } &
     \colhead{$\%$} &
     \colhead{  } &
     \colhead{m$\mu$} &
     \colhead{  } 
}
\startdata
1     &  6.417 $\pm$   0.367&  5.862&  0.487$\pm$   0.044 & 48.08\nl
2     &  6.702 $\pm$   0.058&  5.078&  0.536$\pm$   0.007 & 0.30\nl
4     &  1.529 $\pm$   0.108&  0.570&  0.727$\pm$   0.098 & 6.53\nl
6     &  1.551 $\pm$   0.102&  1.257&  0.518$\pm$   0.070 & 0.11\nl
8     &  7.605 $\pm$   0.194&  1.186&  0.529$\pm$   0.023 & 1.71\nl 
10    &  7.883 $\pm$   0.115&  0.433&  0.441$\pm$   0.009 & 1.01\nl
11    &  4.528 $\pm$   0.057&  0.326&  0.656$\pm$   0.018 & 1.55\nl 
14    &  4.608 $\pm$   0.133&  3.448&  0.551$\pm$   0.036 & 0.17\nl
16    &  5.372 $\pm$   0.012&  1.278&  0.570$\pm$   0.002 & 4.44\nl
19    &  3.725 $\pm$   0.194&  0.874&  0.470$\pm$   0.034 & 1.51\nl
21    &  3.821 $\pm$   0.196&  0.980&  0.530$\pm$   0.048 & 0.04\nl
22    &  1.343 $\pm$   0.069&  1.057&  0.524$\pm$   0.062 & 1.09\nl
23    &  9.417 $\pm$   0.047&  1.038&  0.534$\pm$   0.004 & 0.02\nl
24    &  1.336 $\pm$   0.017&  0.335&  0.579$\pm$   0.013 & 0.54\nl
25    &  9.038 $\pm$   0.075&  3.961&  0.576$\pm$   0.009 & 0.40\nl
27    &  9.430 $\pm$   0.351&  1.327&  0.567$\pm$   0.045 & 0.07\nl
28    &  5.347 $\pm$   0.102&  0.453&  0.620$\pm$   0.022 & 1.49\nl
30    &  7.799 $\pm$   0.073&  0.213&  0.522$\pm$   0.008 & 0.86\nl
32    &  4.814 $\pm$   0.204&  1.334&  0.512$\pm$   0.036 & 0.19\nl
34    &  4.122 $\pm$   0.057&  0.470&  0.595$\pm$   0.019 & 0.21\nl
43    &  6.673 $\pm$   0.026&  0.289&  0.519$\pm$   0.004 & 0.83\nl
46    &  6.777 $\pm$   0.066&  2.158&  0.513$\pm$   0.013 & 0.73\nl
102   &  4.999 $\pm$   0.030&  1.749&  0.530$\pm$   0.005 & 0.13\nl
103   &  1.844 $\pm$   0.170&  3.265&  0.776$\pm$   0.078 & 0.01\nl
104   &  3.520 $\pm$   0.036&  0.880&  0.585$\pm$   0.012 & 0.47\nl
105   &  5.269 $\pm$   0.014&  1.337&  0.612$\pm$   0.006 & 0.12\nl
106   &  3.578 $\pm$   0.046&  3.826&  0.655$\pm$   0.020 & 0.41\nl
107   &  4.354 $\pm$   0.047&  0.570&  0.562$\pm$   0.012 & 0.06\nl
\enddata
\end{planotable}


\end{document}